\documentclass[twocolumn,prl,aps,showpacs,nobalancelastpage,floatfix,preprintnumbers]{revtex4-1}
\usepackage{graphicx}
\usepackage{atlasphysics} \usepackage{amssymb} 
\usepackage{url}
\usepackage{cancel} \usepackage{morefloats}
\usepackage[colorlinks,breaklinks]{hyperref}
\hypersetup{linkcolor=blue,citecolor=blue,filecolor=black,urlcolor=blue}

\newcommand{\pp}{\mbox{$p+p$}}
\newcommand{\pPb}{\mbox{$p$+Pb}}
\newcommand{\pta}{\mbox{$p_{\mathrm{T}}^{\mathrm{a}}$}}
\newcommand{\ptb}{\mbox{$p_{\mathrm{T}}^{\mathrm{b}}$}}
\newcommand{\ptab}{\mbox{$p_{\mathrm{T}}^{\mathrm{a,b}}$}}
\newcommand{\pA}{\mbox{$p$+A}}
\newcommand{\PbPb}{\mbox{Pb+Pb}}
\newcommand{\invmub}{\mbox{$\mu{\rm b}^{-1}$}}
\newcommand{\nchave}{\mbox{$\langle N_{\mathrm{ch}}\rangle$}}
\newcommand{\signch}{\mbox{$\sigma_{\tiny{N_{\mathrm{ch}}}}$}}

\newcommand{\zvtx}{\mbox{$z_{\mathrm{vtx}}$}}

\newcommand{\ETfcal}{\mbox{$\Sigma E_{\mathrm{T}}^{{\scriptscriptstyle \mathrm{Pb}}}$}}

\newcommand{\sqrtsnn}{\mbox{$\sqrt{s_{\mathrm{NN}}}$}}
\newcommand{\Dphi}{\mbox{$\Delta \phi$}}
\newcommand{\Deta}{\mbox{$\Delta \eta$}}

\newcommand{\Yphi}{\mbox{$Y(\Delta\phi)$}}
\newcommand{\Yint}{\mbox{$Y_{\mathrm{int}}$}}
\newcommand{\DelYint}{\mbox{$\Delta Y_{\mathrm{int}}$}}

\usepackage{preprintcover}  
\PreprintCoverPaperTitle{Observation of Associated Near-side and Away-side Long-range
Correlations in $\sqrtsnn =5.02$~\TeV\ Proton--lead Collisions
with the ATLAS Detector}
\PreprintIdNumber{CERN-PH-EP-2012-366}  
\PreprintCoverAbstract{
Two-particle correlations in relative azimuthal angle (\Dphi) and
pseudorapidity (\Deta) are measured in $\sqrtsnn = 5.02$~$\TeV$ 
\pPb\ collisions using the ATLAS detector at the LHC.  
The measurements are performed using approximately 1~\invmub\ of data as a function of transverse momentum ($\pT$) and 
the transverse energy (\ETfcal) summed over $3.1 < \eta < 4.9$ in the direction of
the Pb beam. The correlation function, constructed from charged particles,
exhibits a long-range ($2<|\Delta\eta|<5$) near-side ($\Dphi \sim 0$) correlation that grows 
rapidly with increasing \ETfcal. A long-range away-side ($\Delta \phi \sim\pi$) correlation, 
obtained by subtracting the expected contributions from recoiling dijets and 
other sources estimated using events with small \ETfcal, 
is found to match the near-side correlation in magnitude, 
shape (in $\Delta \eta$ and $\Delta \phi$) and \ETfcal\ dependence.
The resultant \Dphi\ correlation is approximately symmetric about $\pi/2$, 
and is consistent with a dominant $\cos 2\Delta \phi$ modulation for
all \ETfcal\ ranges and particle \pT. 
}
\PreprintJournalName{Phys. Rev. Lett.}

\begin{document}

\title{Observation of Associated Near-side and Away-side Long-range Correlations in $\sqrtsnn =5.02$~\TeV\ Proton--lead Collisions with the ATLAS Detector}
\author{ATLAS Collaboration}

\begin{abstract}
Two-particle correlations in relative azimuthal angle (\Dphi) and
pseudorapidity (\Deta) are measured in $\sqrtsnn = 5.02$~$\TeV$ 
\pPb\ collisions using the ATLAS detector at the LHC.  
The measurements are performed using approximately 1~\invmub\ of data as a function of transverse momentum ($\pT$) and 
the transverse energy (\ETfcal) summed over $3.1 < \eta < 4.9$ in the direction of
the Pb beam. The correlation function, constructed from charged particles,
exhibits a long-range ($2<|\Delta\eta|<5$) near-side ($\Dphi \sim 0$) correlation that grows 
rapidly with increasing \ETfcal. A long-range away-side ($\Delta \phi \sim\pi$) correlation, 
obtained by subtracting the expected contributions from recoiling dijets and 
other sources estimated using events with small \ETfcal, 
is found to match the near-side correlation in magnitude, 
shape (in $\Delta \eta$ and $\Delta \phi$) and \ETfcal\ dependence.
The resultant \Dphi\ correlation is approximately symmetric about $\pi/2$, 
and is consistent with a dominant $\cos 2\Delta \phi$ modulation for
all \ETfcal\ ranges and particle \pT.
\end{abstract}
\maketitle
Proton--nucleus ($\pA$) collisions at the Large Hadron Collider (LHC)
provide both an interesting environment for the study of QCD at high
parton density and important baseline measurements, especially 
for the interpretation of results from the LHC $\PbPb$ program~\cite{Salgado:2011wc}.
In particular, it has been
suggested that $\pPb$ collisions at LHC energies are an important
system for the study of a possible saturation of the growth of parton 
densities at low Bjorken-$x$.

High-multiplicity events provide a rich environment for studying
observables associated with high parton densities in hadronic collisions.
An important tool to probe the physics of these events
is the two-particle correlation function measured in terms of the relative pseudorapidity ($\Delta\eta$) 
and azimuthal angle ($\Delta\phi$) of selected particle pairs, $C(\Delta\eta,\Delta\phi)$.
The first studies of
two-particle correlation functions in the highest-multiplicity \pp\ collisions at the
LHC \cite{Khachatryan:2010gv} showed an enhanced production of pairs of particles at
$\Dphi\sim0$, with the correlation
extending over a wide range in \Deta, a
feature frequently referred to as a ``ridge.'' 
Many of the physics mechanisms proposed to explain the \pp\ ridge,
including multi-parton interactions \cite{Ryskin:2011qh}, parton saturation
\cite{Dumitru:2010iy,Dusling:2012iga,Dusling:2012wy}, and collective expansion of the
final state \cite{Avsar:2010rf},
are also expected to be relevant in $\pPb$ collisions.
A recent measurement by the CMS collaboration~\cite{CMS:2012qk} has demonstrated that a ridge 
is clearly visible over $|\Deta|<4$ in high-multiplicity \pPb\ collisions 
at the LHC.
During final preparation of this Letter, 
the ALICE collaboration submitted a Letter addressing similar physics, 
within the range $|\Deta|<1.8$, with some differences in the analysis technique~\cite{ALICE:2012}.

To provide further insight into the physical origin of 
these long-range correlations, 
this Letter presents \mbox{ATLAS}\ measurements of two-particle
angular correlations over $|\Deta|<5$ in
\pPb\ collisions, based on an 
integrated luminosity of approximately 1~$\invmub$ recorded during a
short run in September 2012.  
The LHC was configured with a 4~\TeV\ proton beam and a
1.57~\TeV\ per-nucleon Pb beam that
together produced collisions with a nucleon--nucleon
center-of-mass energy of $\sqrtsnn = 5.02$~\TeV\ and a rapidity shift of
$-0.47$ relative to the \mbox{ATLAS} rest frame~\footnote{
ATLAS uses a right-handed coordinate system with its origin at the nominal
  interaction point (IP) in the center of the detector and the
  $z$-axis along the beam pipe. The $x$-axis points from the IP to the
  center of the LHC ring, and the $y$-axis points upward. Cylindrical
  coordinates $(r,\phi)$ are used in the transverse plane, $\phi$
  being the azimuthal angle around the beam pipe. For the
  \pPb\ collisions, the incident Pb beam traveled in the $+z$
direction. The pseudorapidity is defined in laboratory coordinates in
terms of the polar angle $\theta$ as $\eta=-\ln\tan(\theta/2)$.
}.

The measurements presented in this Letter are performed using the ATLAS
inner detector (ID), forward calorimeters (FCal), minimum-bias trigger
scintillators (MBTS), and the trigger and data acquisition systems~\cite{Aad:2008zzm}. 
The ID measures charged particles within 
$|\eta|<2.5$ using a combination of silicon
pixel detectors, silicon micro-strip detectors, and a straw-tube
transition radiation tracker, all immersed in a 2~T axial magnetic
field \cite{Aad:2010bx}. The MBTS detect charged particles over $2.1
< |\eta| < 3.9$ using two hodoscopes of 16 counters positioned at $z = \pm
3.6\,$m.  The FCal consists of two sections that cover
$3.1 < |\eta| < 4.9$.
The FCal modules are composed of tungsten and copper absorbers with liquid
argon as the active medium, which together provide 10 interaction
lengths of material. Minimum-bias \pPb\ collisions are selected by a
trigger that requires a signal in at least two MBTS counters.

The \pPb\ events used for this analysis are required to have a
reconstructed vertex containing at least two associated tracks,
with its $z$ position satisfying $|\zvtx|< 150$~mm.  Non-collision
backgrounds and photonuclear interactions are suppressed by
requiring at least one hit in a MBTS counter on each side
of the interaction point, and the difference between times measured 
on the two sides to be less
than 10~ns. Events containing multiple \pPb\ collisions (pileup) are
suppressed  by rejecting events with two reconstructed vertices that 
are separated in $z$ by more than 15\,mm. The residual pileup fraction is
estimated to be $\lesssim 10^{-4}$. About 1.95 million events pass
these event selection criteria. 

Charged particle tracks are reconstructed in the ID using an algorithm 
optimized for \pp\ minimum-bias measurements~\cite{Aad:2010ac}.
In this analysis, the tracks are required to have $\pT>0.3$~\GeV\ and $|\eta|<2.5$,
at least seven hits in the silicon detectors (out of a typical value of 11), and  
a hit in the first pixel layer when one is expected.
In addition, the 
transverse ($d_0$) and
longitudinal ($z_0\sin{\theta}$) 
impact parameters of the tracks measured with
respect to the primary vertex are required to be less than 1.5~mm and 
to satisfy $|d_0/\sigma_{d_0} | < 3$ and
$|z_0\sin{\theta}/\sigma_{z}| <3$, respectively, where $\sigma_{d_0}$ and $\sigma_{z}$ are  
uncertainties on $d_0$ and $z_0\sin{\theta}$ obtained
from the track-fit covariance matrix.

The efficiency, $\epsilon(\pT,\eta)$, for track reconstruction and track selection cuts 
is evaluated using \pPb\ Monte Carlo events produced with the HIJING
event generator~\cite{Wang:1991hta} with a
center-of-mass boost matching the beam conditions. The response of the
detector is simulated using GEANT4~\cite{Agostinelli:2002hh,Aad:2010ah} and the
resulting events are reconstructed with the same algorithms as applied to
the data. The efficiency increases with \pT\ by 6\% 
between 0.3 and 0.5 GeV, and varies only weakly
for $\pT > 0.5$ GeV, where it ranges from 82\% at $\eta = 0$ to 70\% at $|\eta| = 2$ 
and 60\% at $|\eta|> 2.4$.
It is also found to vary by less than 2\% over the range of \ETfcal\ 
observed in the \pPb\ data.

\begin{figure}[!t]
\centering
\includegraphics[width=1\columnwidth]{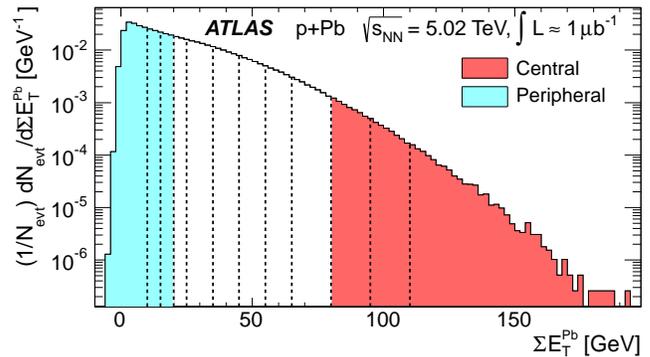}
\caption{\label{fig:cent} Distribution of \ETfcal\ for
  minimum-bias $\pPb$ events. Vertical lines indicate the boundaries
  of the event activity classes. Shaded bands indicate the larger peripheral and central 
intervals having $\ETfcal < 20$~\GeV\ and $\ETfcal > 80$~GeV, respectively.} 
\end{figure}

The two-particle correlation (2PC) analyses are performed in 
different intervals of $\ETfcal$,
the sum of transverse energy measured in the FCal with $3.1<\eta<4.9$
(in the $z$-direction of the lead beam) with no correction for the
difference in response to electrons and hadrons.
The distribution of \ETfcal\ for events passing all selection criteria
is shown in Fig.~\ref{fig:cent}. 
These events are divided into 12 \ETfcal\ intervals (indicated by vertical
lines in Fig.~\ref{fig:cent}) 
to study the variation of 2PC with overall event activity.
Two larger intervals, \mbox{$\ETfcal > 80$~\GeV} and
\mbox{$\ETfcal < 20$~\GeV}, 
containing 2\% and 52\% of the events, respectively, hereafter referred 
to as ``central'' and ``peripheral,'' are used 
for detailed studies of the 2PC at high and low overall event
activity. The quantity $\ETfcal$ instead of charged
particle multiplicity is used to characterize the 
event activity, since the latter is observed to have strong
correlations with the 2PC measurements, particularly for events
selected with low and high multiplicities.
However, for reference, the average (\nchave) and the standard
deviation (\signch) of the efficiency-corrected multiplicity of charged
particles with $\pT>0.4$~\GeV\ and $|\eta|<2.5$ have been 
calculated for each $\ETfcal$ range, yielding $\nchave=150\pm7,
\signch=35\pm2$ for central events and  $\nchave=25\pm1,
\signch=18\pm1$ for peripheral events. 

The 
correlation functions are given~\cite{Adare:2008cqb,Alver:2009id,Aggarwal:2010rf} by:
\begin{eqnarray} 
\label{eq:1}
C(\Dphi,\Deta) = \frac{S(\Dphi,\Deta)}{B(\Dphi,\Deta)}\;,\;\; C(\Dphi) =\frac{S(\Dphi)}{B(\Dphi)},
\end{eqnarray}
where $\Dphi=\phi_a-\phi_b$ and $\Deta=\eta_a-\eta_b$ and $S$ and $B$
represent pair distributions constructed from the same event and from
``mixed events,'' \cite{Adare:2011sc} respectively.
The labels $a$ and $b$ denote the two particles in the pair
(conventionally referred to as ``trigger'' and ``associated'' particles,
respectively~\cite{CMS:2012qk}),
which may be selected from different \pT\ intervals. 
The mixed-event distribution, $B(\Dphi, \Deta)$, that measures
uncorrelated pair yields was constructed by choosing pairs of
particles from different events of similar \zvtx\ and track
multiplicity, to match the effects of detector acceptance, occupancy,
and material on $S(\Dphi,\Deta)$, and of similar \ETfcal. 
The 1D distributions $S(\Dphi)$ and $B(\Dphi)$ are
obtained by integrating $S(\Dphi,\Deta)$ and $B(\Dphi,\Deta)$, 
respectively, over $2<|\Delta\eta|<5$. 
This $|\Delta\eta|$ range is chosen to
focus on the long-range features of the correlation functions.
The normalization of $C(\Dphi,\Deta)$ is chosen
such that the \Dphi-averaged value of $C(\Dphi)$ is unity.
To correct $S(\Dphi,\Deta)$ and $B(\Dphi,\Deta)$ for the inefficiencies, each particle is weighted by 
the inverse of the tracking efficiency.
Remaining detector distortions not accounted for in
the efficiency largely cancel in the same-event to mixed-event ratio.

\begin{figure}[!t]
\centering
\includegraphics[width=1\columnwidth]{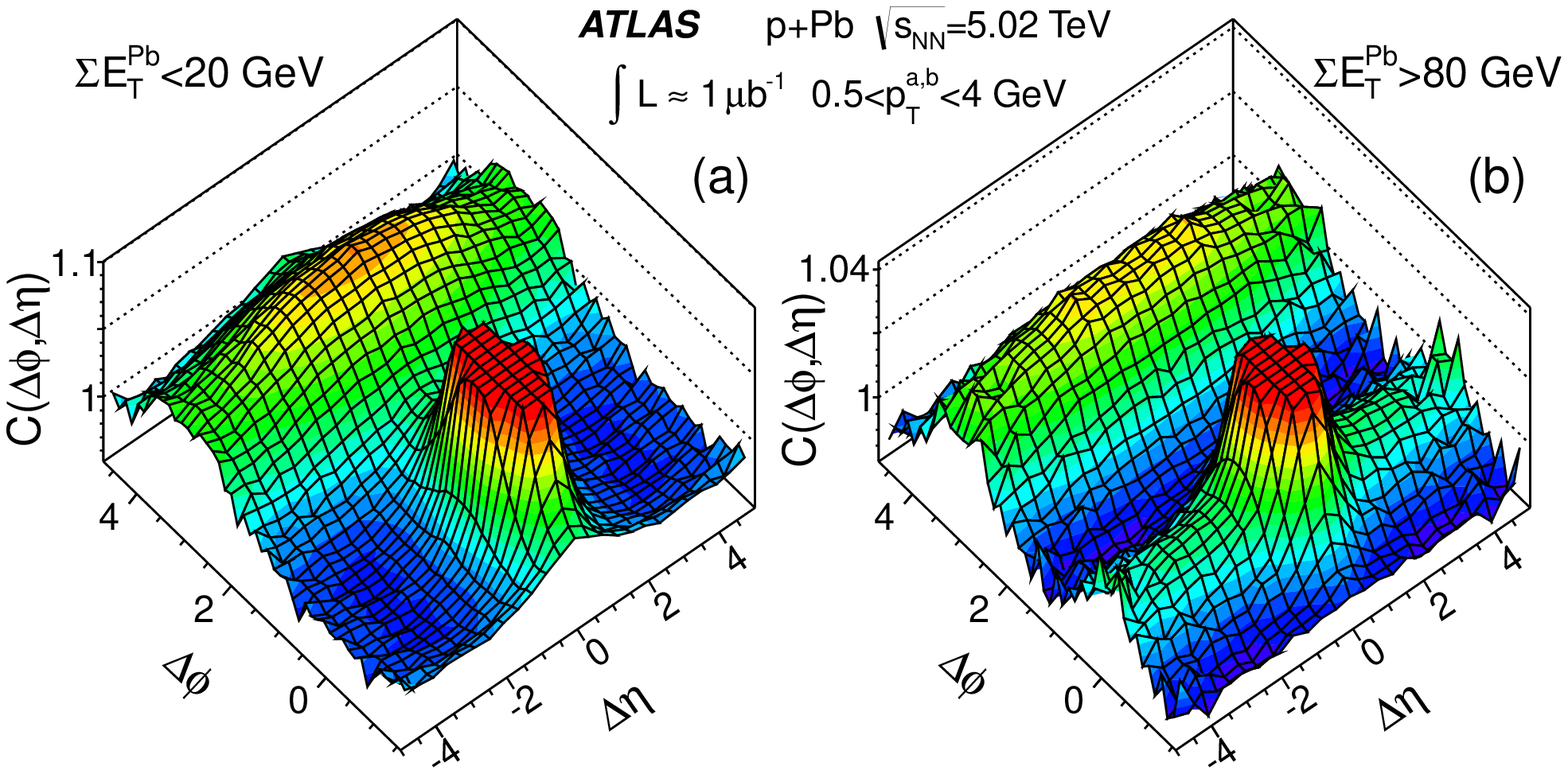}\vspace*{-0.3cm}
\includegraphics[width=1\columnwidth]{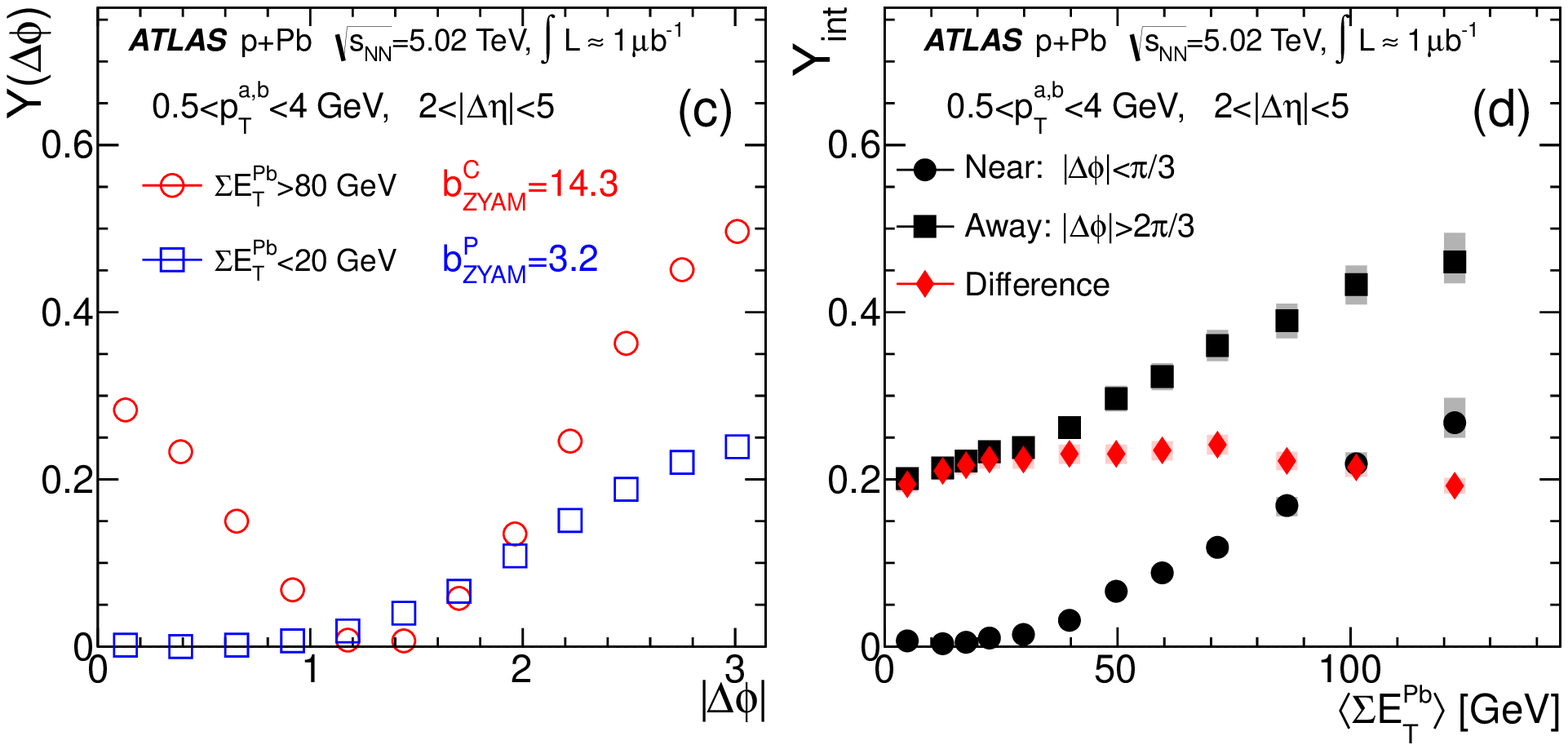}
\caption{\label{fig:method} Two-dimensional correlation functions for 
 (a)  peripheral events and 
 (b)  central events, both with a truncated maximum to suppress the large correlation at $(\Delta\eta,\Delta\phi)=(0,0)$; 
 (c) the per-trigger yield $\Dphi$ distribution together with pedestal levels for peripheral ($b^{\mathrm{P}}_{_{\mathrm{ZYAM}}}$) 
 and central ($b^{\mathrm{C}}_{_{\mathrm{ZYAM}}}$) events, and
 (d) integrated  per-trigger yield as function of \ETfcal\ for pairs in  $2<|\Delta\eta|<5$.
The shaded boxes represent the
systematic uncertainties, and the statistical uncertainties 
are smaller than the symbols.
}
\end{figure}

Examples of 2D correlation functions are shown in Figs.~\ref{fig:method}(a) 
and \ref{fig:method}(b) 
for charged particles with $0.5<\ptab<4$~\GeV\ 
in peripheral and central events.
The correlation function for peripheral events shows a sharp 
peak centered at $(\Dphi,\Deta) = (0,0)$ due to pairs originating from the same jet, Bose-Einstein correlations,
as well as high-$\pT$ resonance decays, and
a broad structure at $\Dphi\sim\pi$ from 
dijets, low-$\pT$ resonances, and momentum conservation that is
collectively referred to as ``recoil'' in the 
remainder of this Letter. 
In the central events, the correlation function reveals a
ridge-like structure at $\Dphi\sim0$ (the ``near-side'') that extends over the full measured \Deta\ range, 
with an amplitude of a few percent.
The distribution at $\Dphi\sim\pi$ (the ``away-side'') is also broadened
relative to peripheral events, consistent with the presence of a
long-range component in addition to that seen in peripheral events.

The strength of the long-range component is quantified by the ``per-trigger yield,'' $Y(\Delta\phi)$,
which measures the average number of particles correlated with each trigger particle,
folded into the $0-\pi$ range~\cite{Adare:2008cqb,Alver:2009id,Aggarwal:2010rf,Khachatryan:2010gv}: 
\begin{eqnarray}
\label{eq:yld}
Y(\Dphi)=\left( \frac{\int B(\Dphi) d\Dphi}{\pi N_{a}} \right) C(\Delta \phi) - b_{_{\mathrm{ZYAM}}},
\end{eqnarray}
where $N_a$ denotes the number of efficiency-weighted trigger particles, and
$b_{_{\mathrm{ZYAM}}}$ represents the pedestal arising from uncorrelated pairs. 
The parameter $b_{_{\mathrm{ZYAM}}}$ is
determined via a zero-yield-at-minimum (ZYAM) 
method~\cite{Ajitanand:2005jj,Adare:2008cqb} in which a second-order polynomial fit
to $C(\Delta \phi)$ is used to find the location of the minimum point, $\Dphi_{_{\mathrm{ZYAM}}}$,
and from this to determine $b_{_{\mathrm{ZYAM}}}$.
The stability of the fit is studied by varying 
the $\Dphi$ fit range.
The uncertainty in $b_{_{\mathrm{ZYAM}}}$ depends on the local
curvature around $\Dphi_{_{\mathrm{ZYAM}}}$, and is estimated to be
0.03\%--0.1\% of the minimum value of $C(\Delta \phi)$. At high \pT\, where the
number of measured counts is low, this uncertainty is of the same order
as the statistical uncertainty.  

The systematic uncertainties due to the tracking efficiency
are found to be negligible for $C(\Dphi)$, since detector effects largely
cancel in the correlation function ratio. However $Y(\Dphi)$
is sensitive to the uncertainty on the tracking efficiency correction 
for the associated particles. 
This uncertainty is estimated by varying
the track quality cuts and the detector material in the simulation,
re-analyzing the data using corresponding Monte Carlo efficiencies and 
evaluating the change in the extracted $Y(\Dphi)$.
The resulting uncertainty on $Y(\Dphi)$ is estimated to be 2.5\% due
to the track selection and 2\%--3\% related to the limited knowledge of detector material. 
The analysis procedure is validated by measuring correlation functions in 
fully simulated HIJING events~\cite{Agostinelli:2002hh,Aad:2010ah} and comparing it to 
the correlations measured using the generated particles.
The agreement is better than 2\% for $C(\Dphi)$
and better than 3\% for $Y(\Dphi)$.

Figure~\ref{fig:method}(c) shows the $Y(\Dphi)$ distributions for
$2<|\Delta\eta|<5$ in 
peripheral and central events separately. The yield for
the peripheral events has an approximate $1-\cos\Dphi$ shape with
an away-side maximum, characteristic of a recoil contribution.
In contrast, the yield in the central events has near-side and away-side peaks 
with the away-side peak having a larger magnitude.
These features are consistent with the onset of a significant $\cos2\Dphi$ 
component in the distribution. To quantify further the properties
of these long-range components, the distributions are integrated over
$|\Dphi|<\pi/3$ and $|\Dphi|>2\pi/3$, and plotted as a function of
\ETfcal\ in Fig.~\ref{fig:method}(d). The near-side yield is close
to 0 for $\ETfcal<20$~\GeV\ and increases with
$\ETfcal$, consistent with the CMS result~\cite{CMS:2012qk}.
The away-side yield shows a similar variation
as a function of $\ETfcal$, except that it starts at a value
significantly above zero, even for events with low $\ETfcal$. The yield
difference between these two regions is found to be approximately independent of
\ETfcal, indicating that the growth in the yield with increasing \ETfcal\ is
the same on the near-side and away-side.

\begin{figure}[!t]
\centering
\includegraphics[width=1\columnwidth]{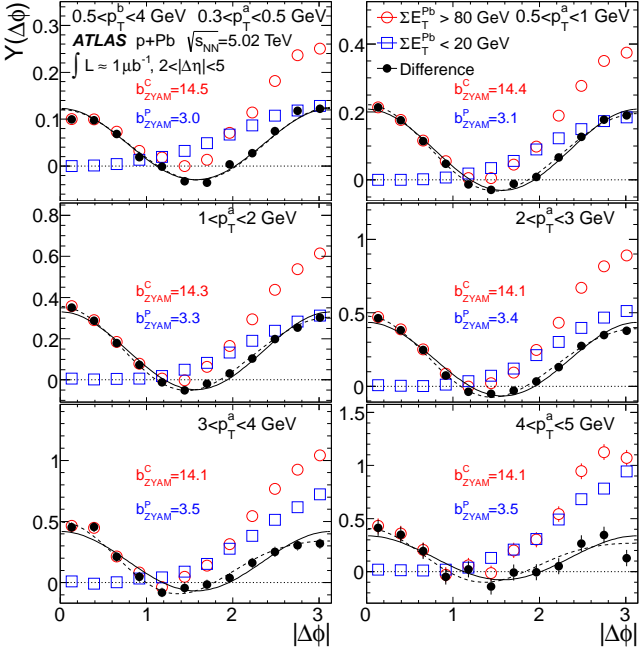}
\caption{\label{fig:cf1d}  Distributions of per-trigger yield in the
peripheral and the central 
event activity classes and their differences (solid 
  symbols), for different ranges of $\pta$ and \mbox{$0.5<\ptb <4$~\GeV},
together with functions
 $a_0+2a_2\cos2\Delta\phi$ (solid line) and $a_0+2a_2\cos2\Delta\phi+2a_3\cos3\Delta\phi$ (dashed line)  obtained via 
a Fourier decomposition (see text). The values for the ZYAM-determined
 pedestal levels are indicated on each panel for peripheral ($b^{\mathrm{P}}_{_{\mathrm{ZYAM}}}$) 
 and central ($b^{\mathrm{C}}_{_{\mathrm{ZYAM}}}$)  \ETfcal\ bins.} 
\end{figure}

To further investigate the connection between the near-side and away-side,
the \Yphi\ distributions for peripheral and central events are shown in Fig.~\ref{fig:cf1d} 
in various \pta\ ranges with
$0.5 < \ptb < 4$~\GeV. Distributions of the difference between central
and peripheral yields, $\Delta \Yphi$, are also shown in this
Figure. This difference is observed to be nearly symmetric around
$\Dphi = \pi/2$. To illustrate this symmetry, the $\Delta
\Yphi$ distributions in Fig.~\ref{fig:cf1d} are overlaid with 
functions $a_0 + 2a_2 \cos{2\Dphi}$ and $a_0 + 2a_2 \cos{2\Dphi}+ 2a_3 \cos{3\Dphi}$, with the coefficients 
calculated as \mbox{$a_n=\langle\Delta\Yphi\cos n\Dphi\rangle$}. 
Using only the $a_0$ and $a_2$ terms describes the $\Delta
Y$ distributions reasonably well, indicating that the long-range component of the
two-particle correlations can be approximately described by a recoil
contribution plus a $\Dphi$-symmetric component. The inclusion of the $a_3$ term improves slightly the agreement with the data.

\begin{figure}[!t]
\centering
\includegraphics[width=0.90\columnwidth]{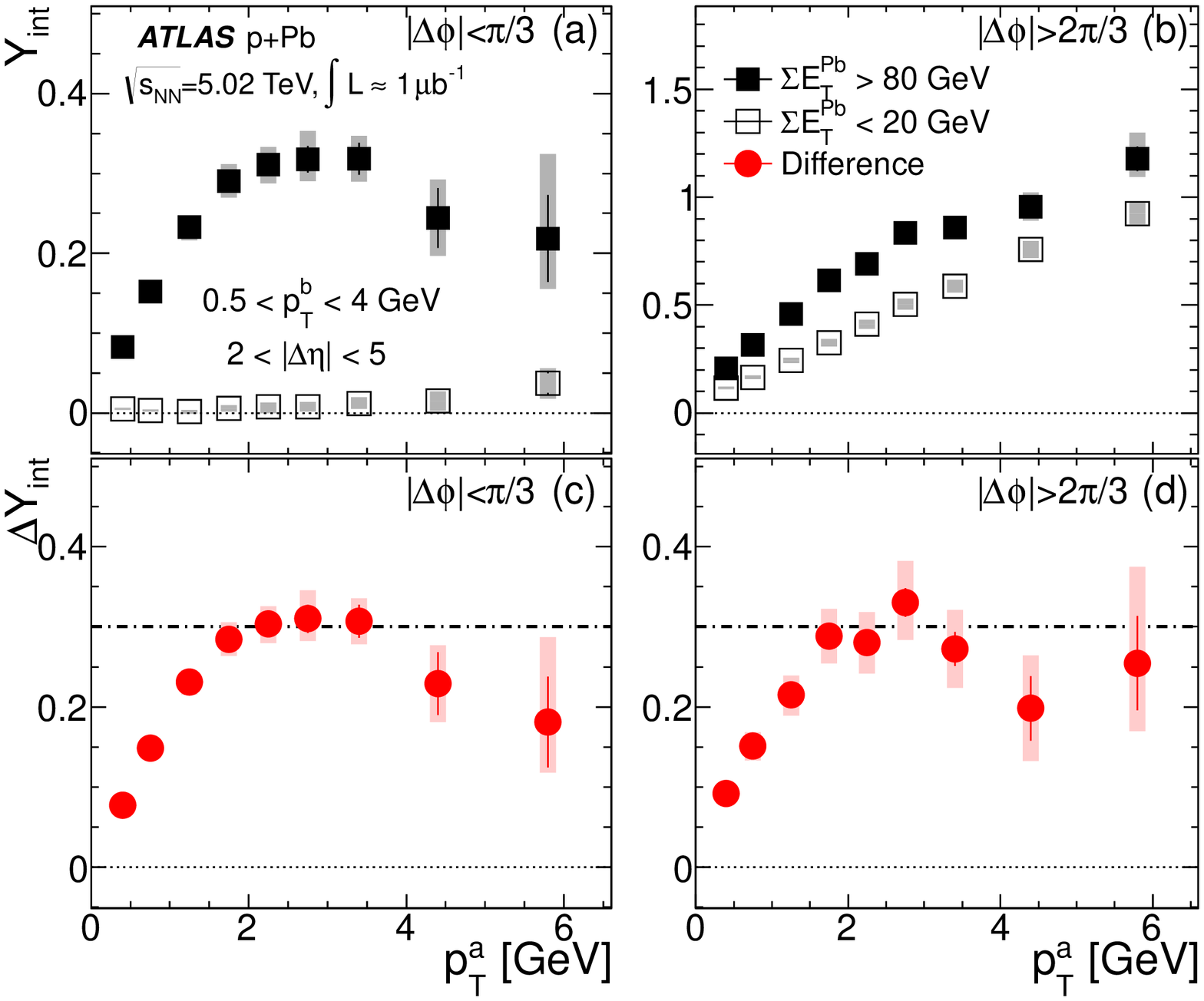}\\
\includegraphics[width=0.45\columnwidth]{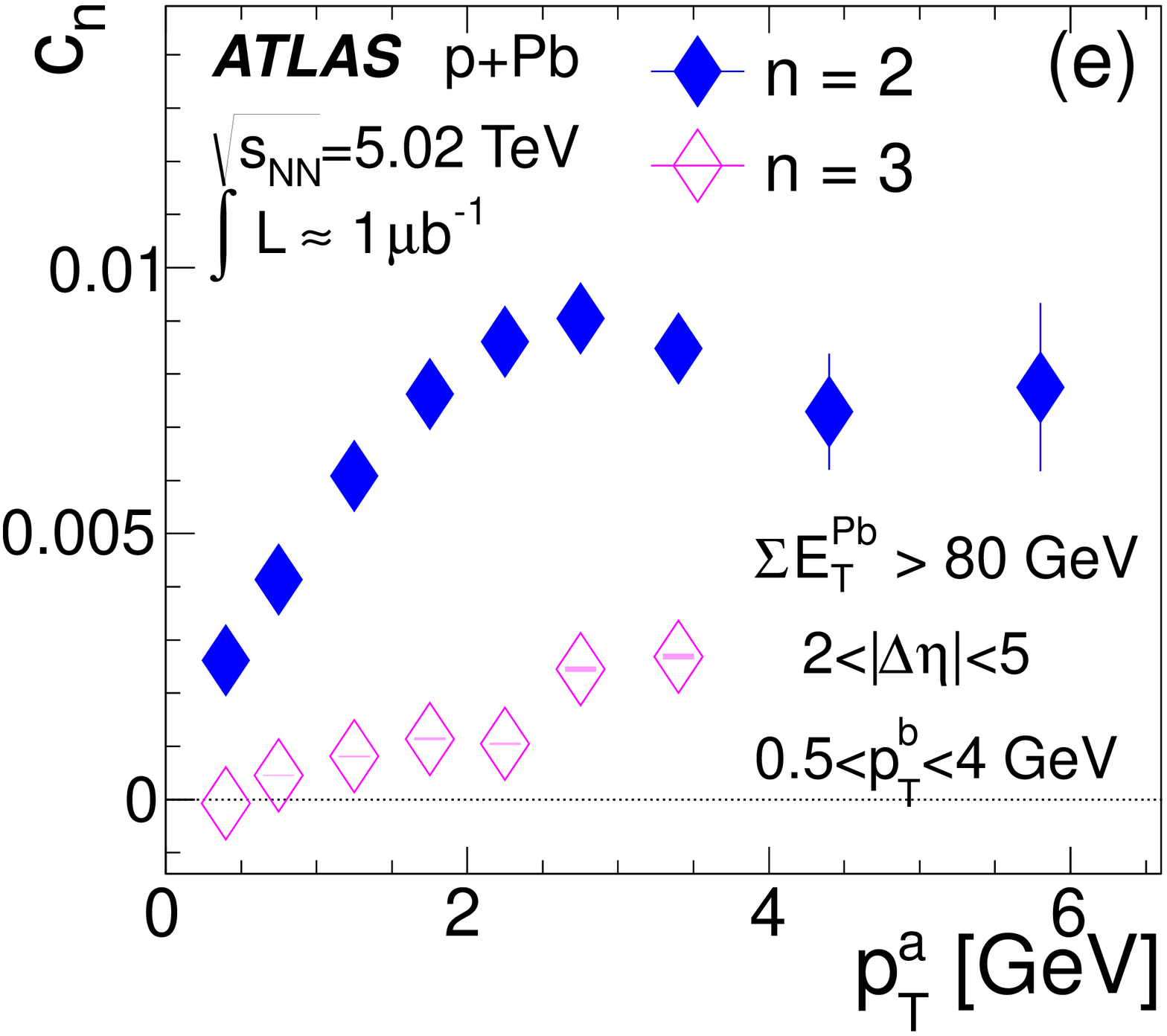}\includegraphics[width=0.45\columnwidth]{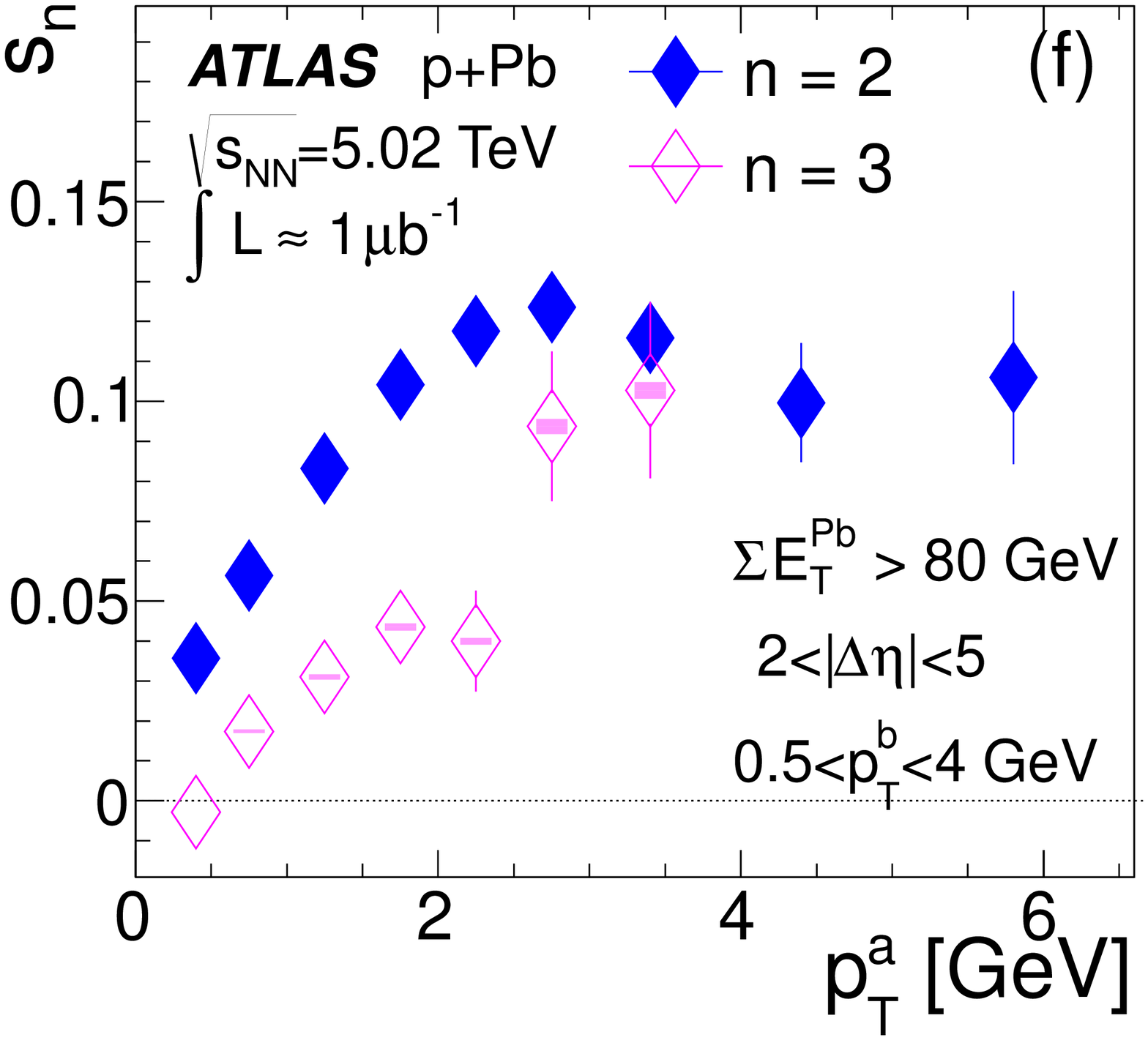}
\caption{\label{fig:res} Integrated per-trigger yields, \Yint (see
  text), vs $\pta$ 
  for $0.5<\ptb<4$~\GeV\, in peripheral and central events,
on the (a) near-side and (b) away-side.  The panels
  (c) and (d) show the difference, \DelYint. Panels (e) and (f) show the
  $\pT$ dependence of $c_n$ and $s_n$ for n=2,3, respectively. 
 The error bars and shaded boxes represent the
  statistical and systematic uncertainties, respectively.} 
\end{figure}

The near-side and away-side yields integrated over $|\Dphi|<\pi/3$ and
$|\Dphi|>2\pi/3$, respectively (\Yint), and the differences between
those integrated 
yields in central and peripheral events (\DelYint) are shown in
Fig.~\ref{fig:res} as a 
function of \pta. The yields are shown separately for the two \ETfcal\
ranges in panels (a)--(b) and the differences are shown in panels
(c)--(d). Qualitatively, the differences have a similar \pta\ dependence and 
magnitude on the near-side and away-side; they rise with \pta\ and reach a maximum
around 3--4~\GeV.
This pattern is visible for the
near-side even before subtraction, as shown in panel (a), but is less
evident in the unsubtracted away-side due to the dominant contribution
of the recoil component. A similar dependence is observed for 
long-range correlations in
\PbPb\ collisions at approximately the same
$\pT$~\cite{Chatrchyan:2012wg,ATLAS:2012at}.

The relative amplitude of the $\cos{n\Dphi}$ modulation of $\Delta
\Yphi$, $c_n$, for $n=2,3$ can be estimated using $a_n$, and the
extracted value of $b_{_{\mathrm{ZYAM}}}$ for central events:
\begin{equation}
c_n = a_n/(b^{{\mathrm{C}}}_{_{\mathrm{ZYAM}}} + a_0).
\label{eq:ctwo}
\end{equation}
Figure~\ref{fig:res}(e) shows $c_2$ and $c_3$ as a function of
\pta\ for $0.5<\ptb<4$~\GeV. The value of $c_2$ is much larger than $c_3$ and exhibits a
behavior similar to $\Delta\Yphi$ at the near-side and away-side.
Using the techniques discussed in Ref.~\cite{ATLAS:2012at},
$c_n$ can be converted into an estimate of $s_n$, 
the average $n^{\mathrm th}$ Fourier coefficient of the event-by-event single-particle $\phi$ distribution,
by assuming the factorization relation $c_n(\pta,\ptb)=s_n(\pta)s_n(\ptb)$.
From this, $s_n(\pta)$ is calculated as $s_n(\pta)=
c_n(\pta,\ptb)/\sqrt{c_n(\ptb,\ptb)}$, where  
$c_n(\ptb,\ptb)$ is obtained from
Eq.~(\ref{eq:ctwo}) using the $a_n$ extracted from the 
difference between the central and peripheral data shown in Fig.~\ref{fig:method}(c). 
The $s_2(\pta)$ values obtained this way exceed
0.1 at $\pT \sim$ 2--4~\GeV, as shown in
Fig.~\ref{fig:res}(f). The $s_3(\pta)$ values are
smaller than $s_2(\pta)$ over the measured $\pT$ range.
The factorization relation used to compute
$s_2(\pta)$ is found to be valid within 
10\%--20\% when selecting different sub-ranges of $\ptb$ 
within 0.5--4~\GeV, while the precision of $s_3(\pta)$ data does not
allow a quantitative test of the factorization. The analysis is also repeated for 
correlation functions separately constructed from like-sign pairs 
and unlike-sign pairs, 
and the resulting $c_n$ and $s_n$ coefficients are found
to be consistent within their statistical and systematic uncertainties. 

In summary, ATLAS has measured two-particle correlation functions in
$\sqrtsnn =5.02$~\TeV\ \pPb\ collisions in different intervals of
\ETfcal\ over $2<|\Delta \eta|<5$.
An away-side contribution is observed that grows
rapidly with increasing \ETfcal\ and which matches many
essential features of the near-side ridge observed here, as well as
in previous high-multiplicity $\pp$, $\pPb$ and $\PbPb$ data at the LHC. Thus, while the 
ridge in \pp\ and \pPb\ collisions has been characterized as a
near-side phenomenon, these results show that it 
has both near-side and away-side components that are symmetric around
$\Dphi \sim \pi/2$, with a \Dphi\ dependence that is approximately
described by a $\cos2\Delta \phi$ modulation.
A Fourier decomposition of the correlation function, $C(\Dphi)$, 
yields a pair $\cos2\Delta \phi$ amplitude of about 0.01 at $\pT \sim 3$~\GeV,
corresponding to a single-particle amplitude of about 0.1. 
Similar findings are obtained independently by the ALICE collaboration~\cite{ALICE:2012}, 
albeit over a more restricted phase space ($|\Deta|<1.8$ and $\pT<$2--4 GeV).
The two results are found to be consistent within this common region.

Some of the features of the data, including the presence of an away-side component, 
are qualitatively predicted in the Color Glass  
Condensate approach~\cite{Dusling:2012wy} which models saturation of
the parton distribution in the Pb nucleus. 
The estimated amplitudes of the modulation on the single-particle level
are also found to be comparable in magnitude and $\pT$ dependence
to similar modulations observed in heavy-ion collisions,
commonly attributed to collective expansion of the hot, dense matter~\cite{ATLAS:2012at}.
Thus, although the original motivation for this work was to study the
possible effects of high parton density in the initial state of $\pPb$
collisions, the results presented here are also consistent with
contributions of final-state collective effects in high-multiplicity events~\cite{Bozek:2011if,Bozek:2012gr}.
\section*{Acknowledgments}

We thank CERN for the very successful operation of the LHC, as well as the
support staff from our institutions without whom ATLAS could not be
operated efficiently.

We acknowledge the support of ANPCyT, Argentina; YerPhI, Armenia; ARC,
Australia; BMWF, Austria; ANAS, Azerbaijan; SSTC, Belarus; CNPq and FAPESP,
Brazil; NSERC, NRC and CFI, Canada; CERN; CONICYT, Chile; CAS, MOST and NSFC,
China; COLCIENCIAS, Colombia; MSMT CR, MPO CR and VSC CR, Czech Republic;
DNRF, DNSRC and Lundbeck Foundation, Denmark; EPLANET and ERC, European Union;
IN2P3-CNRS, CEA-DSM/IRFU, France; GNAS, Georgia; BMBF, DFG, HGF, MPG and AvH
Foundation, Germany; GSRT, Greece; ISF, MINERVA, GIF, DIP and Benoziyo Center,
Israel; INFN, Italy; MEXT and JSPS, Japan; CNRST, Morocco; FOM and NWO,
Netherlands; RCN, Norway; MNiSW, Poland; GRICES and FCT, Portugal; MERYS
(MECTS), Romania; MES of Russia and ROSATOM, Russian Federation; JINR; MSTD,
Serbia; MSSR, Slovakia; ARRS and MVZT, Slovenia; DST/NRF, South Africa;
MICINN, Spain; SRC and Wallenberg Foundation, Sweden; SER, SNSF and Cantons of
Bern and Geneva, Switzerland; NSC, Taiwan; TAEK, Turkey; STFC, the Royal
Society and Leverhulme Trust, United Kingdom; DOE and NSF, United States of
America.

The crucial computing support from all WLCG partners is acknowledged
gratefully, in particular from CERN and the ATLAS Tier-1 facilities at
TRIUMF (Canada), NDGF (Denmark, Norway, Sweden), CC-IN2P3 (France),
KIT/GridKA (Germany), INFN-CNAF (Italy), NL-T1 (Netherlands), PIC (Spain),
ASGC (Taiwan), RAL (UK) and BNL (USA) and in the Tier-2 facilities
worldwide.

\bibliography{pPbRidgePRL}
\bibliographystyle{apsrev4-1}
\clearpage
\onecolumngrid
\section*{{\Large ATLAS Collaboration}}
\label{app:ATLASColl}

\bigskip

G.~Aad$^{\rm 48}$,
T.~Abajyan$^{\rm 21}$,
B.~Abbott$^{\rm 111}$,
J.~Abdallah$^{\rm 12}$,
S.~Abdel~Khalek$^{\rm 115}$,
A.A.~Abdelalim$^{\rm 49}$,
O.~Abdinov$^{\rm 11}$,
R.~Aben$^{\rm 105}$,
B.~Abi$^{\rm 112}$,
M.~Abolins$^{\rm 88}$
% ...
\bigskip

% ...
$^{11}$ Institute of Physics, Azerbaijan Academy of Sciences, Baku, Azerbaijan\\
$^{12}$ Institut de F\'{i}sica d'Altes Energies and Departament de F\'{i}sica de la Universitat Aut\`{o}noma de Barcelona and ICREA, Barcelona, Spain\\
% ...
$^{21}$ Physikalisches Institut, University of Bonn, Bonn, Germany\\
% ...
$^{48}$ Fakult\"{a}t f\"{u}r Mathematik und Physik, Albert-Ludwigs-Universit\"{a}t, Freiburg, Germany\\
$^{49}$ Section de Physique, Universit\'{e} de Gen\`{e}ve, Geneva, Switzerland\\
% ...
$^{88}$ Department of Physics and Astronomy, Michigan State University, East Lansing MI, United States of America\\
% ...
$^{105}$ Nikhef National Institute for Subatomic Physics and University of Amsterdam, Amsterdam, Netherlands\\
% ...
$^{111}$ Homer L. Dodge Department of Physics and Astronomy, University of Oklahoma, Norman OK, United States of America\\
$^{112}$ Department of Physics, Oklahoma State University, Stillwater OK, United States of America\\
% ...
$^{115}$ LAL, Universit\'{e} Paris-Sud and CNRS/IN2P3, Orsay, France\\
% ...

\end{document}